# Time Until Neuron Death After Initial Puncture From an Amyloid-Beta Oligomer

**Tanner Dalton Davis Horton**
*Senior Project thesis paper, May 2013- May 2014*
*Yreka High School*
*Advisor and editor: Austin Sendek*

**Abstract:** Hardy and Higgins first proposed the amyloid cascade hypothesis in 1992, stating that the decrease in neuronal function observed in Alzheimer's Disease (AD) is due to a process initiated by the oligomerization of amyloid beta peptide. One hypothesis states that toxicity arises from the aggregation of amyloid beta into a pore structure, which can then puncture the brain cell membrane; this allows toxic $Ca^{2+}$ ions to flood through the pore, causing eventual cell death. In 2007, neurobiologist Ruth Nussinov calculated the three pore sizes most likely to occur within the brain. Based on her findings, we constructed a method to determine the time it takes for a cell to die after the cell is punctured by the pore. Our findings have shown that cell death occurs within one second after the Aβ oligomer makes contact with the cell. We believe this is important because instant cell death has been one criticism of Nussinov's model, and we have calculated a concrete time value for that criticism. We identify two potential deficiencies with our model that could be improved: first, we treat $Ca^{2+}$ in our model as an ideal gas, which it is not; second, we assume that the pores are static (i.e. constantly open), while recent developments suggest they may open and close dynamically.

# Table of Contents

# I. Introduction

**Alzheimer's disease**
Alzheimer's disease (AD) is a degenerative brain disease which can lead to memory loss, a decrease in basic motor and thinking skills, as well as the loss of ability to carry out the most basic tasks. One of the most peculiar facts about AD is that scientists have been unable to determine an exact cause. Scientists agree that lifestyle, genetics, as well as environmental factors can attribute to the development of AD.[1] Due to the complexity of the disease, no cure has been found for it. Several hypotheses currently exist that shed light as to the gradual development of AD. Better understanding of these hypotheses, such as the amyloid cascade hypothesis, can lead to potential cures and preventive treatments in the future.

**Amyloid cascade hypothesis**
This hypothesis proposes that AD is caused due to the oligomerization of the amyloid-beta peptide. These peptides are the result of secretases splitting apart amyloid precursor proteins (APP). The Aβ peptides then group together to form an oligomer. Post-mortem observations of brain tissue from patients with Alzheimer's disease have shown an increased level of neuritic plaques. Aβ oligomers are the main component of these plaques.[2]

It is not clear exactly how the Aβ oligomerization causes cell death. Toxicity could arise from a number of different sources: the process of oligomerization, the oligomerized products themselves, or some off-pathway mechanism initiated by the aggregation (such as degradation of tau protein). One hypothesis states that the oligomers form a ring structure which allows them to puncture brain cells, creating a pore enabling the process of diffusion of toxic species to occur. In particular, $Ca^{2+}$ ions will move from a high concentration outside of the cell through the pore into a concentration $1.3 \times 10^3$ times smaller. The increased presence of this toxin within the brain cell is what is believed to damage the cell, thereby decreasing neuron function.

As the number of affected neurons begin to die, it is then that the common symptoms of Alzheimer's transpire. Several attempts have been made to prevent this occurrence, including limiting amyloid-beta production and preventing aggregation. Both of these possible cures failed during clinical trials.[3]

## Nussinov Models

By employing computational models, Nussinov et al have demonstrated that Aβ monomers can form stable ring structures which can then function as pores. The three most stable ring structures have the following diameters: 2.2 nm, 2.5 nm, and 2.7 nm[4]. These simulated sizes are in agreement with experimental observations.

The atomic radius of calcium is ~200 pm, which means a 2.5 nm diameter pore has a diffusion area of approximately 150 times that of the cross-sectional area of a calcium atom. This large difference in length scales means that the calcium can be treated as an ideal gas which will flow through the pore.

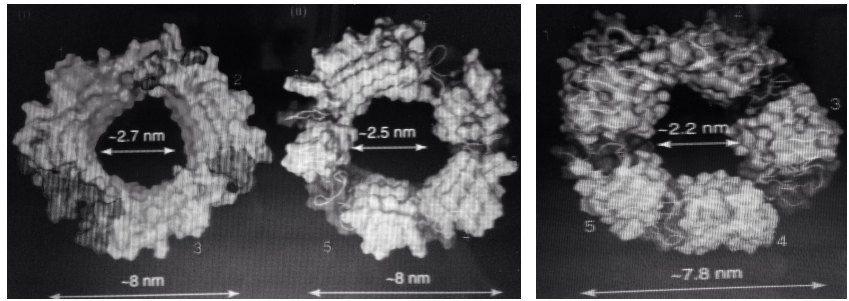

## Our hypothesis

Alzheimer's disease can take several months to several years to fully develop[5]. We hypothesize that this "onset time" between first Aβ aggregation and severe AD is a result of the slow diffusion of calcium through Aβ pores into the cells until reaching a lethal concentration. Due to this prolonged progression, we predict that the time required to reach lethal calcium toxicity will be on the magnitude of years, around $10^7$ to $10^8$ seconds.

# II. Model

## Our equation

The foundation for our model is based off of the Ideal Gas Law. By treating calcium inside of the brain like an ideal gas, we are able to draw comparisons between neuronal calcium and the diffusion of gas atoms across a concentration gradient. We employ statistical mechanics models to develop this equation into one which can calculate the concentration of calcium inside the cell per time; knowing the lethal concentration, we can calculate the time it takes for a cell to die after being punctured.

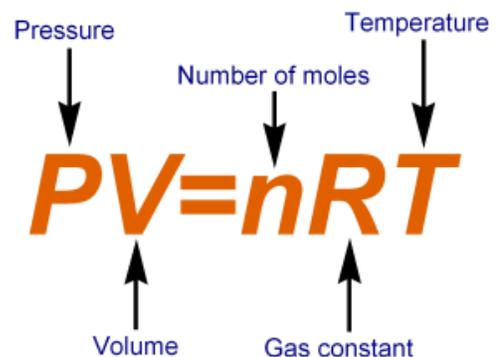

After bringing in initial conditions of the brain as well as several other factors, our final equation came to be:

$$t = \frac{-N_A V_C}{\alpha} \ln\left[\frac{M_{lethal} - \Delta M}{2M_{in} - M_{out}}\right], \text{ where } \alpha = \frac{A_p RT}{2\sqrt{3kTm}}$$

| Variable | Units | Actual value |
| --- | --- | --- |
| $N_A$, Avogadro's constant | particles per mole | 6.022e23 |
| $V_C$, volume of the cell | meters³ | 5.964e-5 |
| $M_{lethal}$, toxic Ca²⁺ limit | moles | 5e-6 |
| $M_{in}$, Ca²⁺ molarity inside of the cell | moles | 1.5e-7 |
| $M_{out}$, Ca²⁺ molarity outside of the cell | moles | 2e-3 |
| $A_p$, area of the pore | nanometers² ⇒ meters² | π (Radius of pore)² |
| $R$, gas constant | Joules per Kelvin mole | 8.314 |
| $T$, temperature in the brain | Kelvin | 310 |
| $k$, Boltzmann's constant | Joules per Kelvin | 1.3806e-23 |
| $m$, mass of a Ca²⁺ ion | kilograms | 6.65e-26 |

For the full derivation of our equation, refer to the appendix.

# III. Results

**Final times**

| Pore Diameter (nm) | Time (s) |
| --- | --- |
| 2.2 | 0.0514 |
| 2.5 | 0.0398 |
| 2.7 | 0.0341 |

# IV. Discussion

**Implications for the Nussinov model**
A common criticism of Nussinov's pore model is that the relatively large pore size means that calcium diffusion will be so fast that the cells will die almost immediately upon puncture. However, immediate cell death upon puncture seems to be at odds with experimental evidence, suggesting that the Nussinov pore model is flawed.

We set out to use this model to show that the time required to reach lethal calcium toxicity in punctured neurons is on the timescale of years, given the years-long onset time of AD. However, we have instead found that our calculations lend quantitative support for such criticisms of the model. If we assume experimental evidence is correct, then the Nussinov model predicts unrealistic outcomes. There must be an incorrect assumption somewhere in the Nussinov model.

There are a few ways in which AB pore hypothesis may be saved. First, the pore may take on a different shape than a simple cylinder. It has been suggested that the pores may be hourglass shaped -- having a smaller radius in the middle than on the ends. This would decrease the surface area open to calcium diffusion. Second, the pores may be dynamic rather than static. Our model assumes the pores are constantly open, but there is reason to believe that the pores can open and close selectively after puncturing the cell. If the pores are only open for microseconds at a time, then clearly the amount of time needed for cell death will increase.

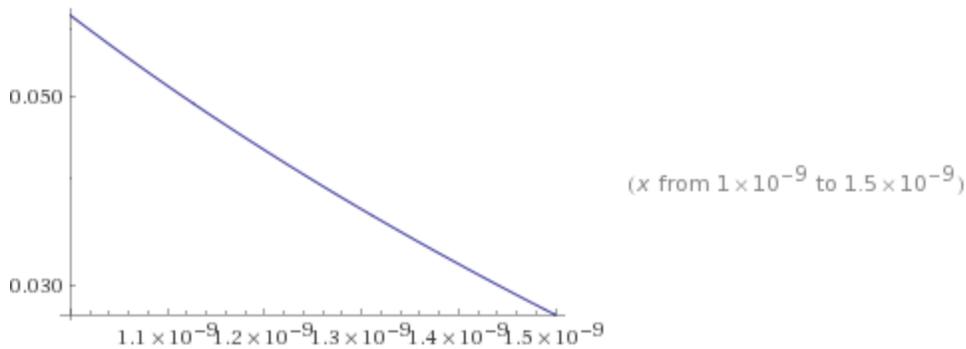

If the pore is truly static and cylinder shaped, the radius must be much smaller than 1.1 nm to generate a ~1 year cell death time. Above we have plotted cell death time vs. pore radius; in order to have a cell death time of 1 year, a small pore size of 4.44e-14 meters is required. This is about 0.02% of the atomic radius of a calcium ion, which renders this situation impossible; a calcium ion could never diffuse through such a pore. This suggests there is a deeper flaw in the model.

**Future improvements on our model**
The first flaw in our model lies in the fact that we treat the calcium as if it is an ideal gas. An ideal gas is, by definition, a set of randomly moving, non-interacting particles; calcium ions however exhibit electrostatic interactions due to the ionic charge of +2. New terms accounting for this non-ideality could be introduced to our equation in order to make the assumption unnecessary. Additional errors may be introduced by the fact that the pore radius is only about five times larger than calcium's atomic radius. Typically, ideal gas models are only valid when the length scales involved are much bigger than the atomic radius of the gas. To make the model more realistic, it should be adjusted to account for the comparable pore and ionic sizes.

We may also need to reconsider the assumption that the calcium concentration outside the cell is constant in time. With such a large flux of calcium into the cell over such a short period of time, it's possible that the amount of calcium around the cell may become depleted and diffusion into the cell will slow down.

The model could also be improved by considering the aforementioned changes to the Nussinov model: dynamic pores and differing pore geometries. Last, our model only explores the effect of one puncture at a time; the possibility of multiple punctures at once should be taken into consideration.

# V. Appendix

## Derivation of our equation

The ideal gas law can be rewritten as PV=NkT. We first find the average pressure exerted on the cell by each colliding atom by considering the force per area exerted on the membrane by each collision. Taking advantage of Newton's second and third laws, we write:

$$\overline{P} = \frac{\overline{F_{x,\text{ on inside of cell}}}}{A} = \frac{-\overline{F_{x,\text{ on molecules}}}}{A} = -\frac{m\overline{\left(\frac{\Delta v_x}{\Delta t}\right)}}{A}$$

Assuming the collision is elastic, throughout this interval, the change in the velocity of the molecule is given by:

$$\Delta v_x = (v_x,\ \text{final}) - (v_x,\ \text{initial}) = -2v_x$$

Combining the last three equations produces the average pressure:

$$\overline{P} = \frac{2mv_x}{A\Delta t}$$

This is an expression for the average pressure exerted per atom; to find the total pressure, we sum over every atom by multiplying the right hand side by N, the number of atoms. $\overline{P}$ goes to P, and $v_x$ goes to $\overline{v_x}$, the average atomic velocity.

For the average velocity, we employ the equipartition theorem, which says the kinetic energy stored in each dimension is $\frac{kT}{2}$. For an atom moving in three dimensions, the average kinetic energy is:

$$\overline{\tfrac{1}{2}mv^2} = \tfrac{1}{2}m\overline{\left(v_x^2 + v_y^2 + v_z^2\right)} = \tfrac{1}{2}kT + \tfrac{1}{2}kT + \tfrac{1}{2}kT = \tfrac{3}{2}kT$$

Solving for $\overline{v^2}$ and then taking the square-root gives us $v_{rms}$, or the root-mean-square velocity:

$$v_r = \sqrt{\frac{3kT}{m}} = v_{rms}$$

Using this for the average atomic velocity, we arrive at the below expression for the calcium pressure on an area *A* of the cell membrane as a function of calcium's mass, velocity, and concentration:

$$P = \frac{2mNv_r}{A\Delta t}$$

In a living cell, there are many different atoms and molecules floating around and exerting pressure on the cell membrane. We only want to consider the effect of calcium. Therefore, we replace P with the osmotic pressure, $p_{osmotic}$, which we can calculate using the Morse equation:

$$P = p_{osmotic} = RT\Delta M$$

where $\Delta M$ is the change in calcium concentration across the membrane.

Since there is a large number of calcium ions throughout the body, we consider the number of calcium ions immediately outside the cell, N, to be constant in time. Putting this together, we now have an expression for the calcium concentration difference across an area A of the cell membrane as a function of the particle number outside of A:

$$RT\Delta M = \frac{2mN\sqrt{\frac{3kT}{m}}}{A\Delta t}$$

$$N = \frac{ART\Delta M \Delta t}{2\sqrt{3kTm}}$$

Now, we imagine that we insert a pore of area A on the cell membrane which allows calcium to flow freely into and out of the cell. For any given calcium ion that makes contact with the cell membrane, the probability that it hits the pore (and enters the cell) is simply the ratio of the pore area to the full cell area:

$$Prob(Ca^{2+} \text{ enters}) = \frac{A_{pore}}{A_{cell}},$$

so the number of calcium ions entering the cell as a function of the number of calcium ions immediately outside the cell is:

$$\Delta N = \frac{A_{pore}}{A_{cell}} N_{out}$$
$$= \frac{A_{pore}}{A_{cell}} x \frac{PA_{cell}\Delta t}{2\sqrt{3kTm}}$$
$$= \frac{RT\Delta M A_{pore}\Delta t}{2\sqrt{3kTm}}$$

Dividing through by $\Delta t$ and taking the limit as $\Delta t$ goes to zero gives an expression for the rate of change of calcium ions inside the cell, $\frac{dN}{dt}$:

$$\lim_{\Delta t \to 0} \frac{\Delta N}{\Delta t} = \frac{dN}{dt} = \frac{RT\Delta M A_{pore}}{2\sqrt{3kTm}}$$

Note that $\Delta M$, the concentration difference across the cell membrane, changes in time as calcium flows into the cell. $\Delta M$ can be rewritten as the difference between the initial

(constant) calcium concentration outside the cell $M_{out}$ and the sum of the initial internal concentration plus the differential increase over time, $\delta M(t)$:

$$\Delta M = M_{out} - [M_{in}(t=0) + \delta M(t)]$$
$$= M_{out} - [M_{in}(t=0) + N_{in}(t)(\tfrac{1}{N_A V_c})]$$

Where $N_{in}$ is the number of calcium ions in the cell, $N_A$ is Avogadro's number and $V_C$ is the volume of the cell. Plugging this into the formula for $\tfrac{dN}{dt}$, we can separate out the time dependence:

$$\tfrac{dN}{dt} = \tfrac{A_p RT}{2\sqrt{3kTm}} \left[ M_{out} - M_{in}(t=0) - \tfrac{N_{in}}{N_A V_C} \right]$$
$$= \tfrac{A_p RT}{2\sqrt{3kTm}} [M_{out} - M_{in}(t=0)] - \tfrac{A_p RT}{2\sqrt{3kTm}} \left[ \tfrac{N_{in}}{N_A V_C} \right]$$

Now, let $\alpha = \tfrac{A_p RT}{2\sqrt{3kTm}}$, and we can rewrite the expression:

$$\tfrac{dN}{dt} = \alpha \Delta M_o - \tfrac{\alpha N_{in}}{N_A V_C}$$

This is a first-order linear differential equation,

$$\tfrac{dN}{dt} + \tfrac{\alpha N_{in}}{N_A V_C} = \alpha \Delta M_o$$

which has the general solution:

$$N(t) = \tfrac{1}{u(t)} \left[ \int u(t) \alpha \Delta M_o \, dt + C \right], \text{ where } u(t) = e^{\int \tfrac{\alpha}{N_A V_C} dt} = e^{\tfrac{\alpha t}{N_A V_C}}$$

$$= e^{\tfrac{-\alpha t}{N_A V_C}} \left[ \alpha \Delta M_o \int e^{\tfrac{\alpha t}{N_A V_C}} dt + C \right]$$

$$= \alpha \Delta M_o \exp(-\tfrac{\alpha t}{N_A V_C}) \left[ \tfrac{N_A V_C}{\alpha} e^{\tfrac{\alpha t}{N_A V_C}} + C \right]$$

To solve for the integration constant C we use our known initial conditions: when t = 0, $N = M_{o,in} N_A V_C$:

$$N(t=0) = M_{o,in} N_A V_C = \alpha \Delta M_o \left[ \tfrac{N_A V_C}{\alpha} + C \right]$$
$$= \Delta M_o N_A V_C + \alpha \Delta M_o C$$
$$= (M_{out} - M_{o,in} N_A V_C) N_A V_C + \alpha (M_{out} - M_{o,in}) C$$

$$2M_{o,in}N_A V_C = M_{out}N_A V_C + \alpha(M_{out} - M_{o,in})C$$

Thus,

$$\begin{aligned}C &= \frac{2M_{o,in}N_A V_C - M_{out}N_A V_C}{\alpha(M_{out}-M_{o,in})} \\ &= \frac{N_A V_C(2M_{o,in}-M_{out})}{\alpha(M_{out}-M_{o,in})} \\ &= \frac{N_A V_C(2M_{o,in}-M_{out})}{\alpha \Delta M_o}\end{aligned}$$

Plugging in C gives the following equation:

$$\begin{aligned}N(t) &= \alpha \Delta M_o \frac{N_A V_C}{\alpha} + \alpha \Delta M_o e^{\frac{-\alpha t}{N_A V_C}} \left[ \frac{N_A V_C(2M_{o,in}-M_{out})}{\alpha \Delta M_o} \right] \\ &= N_A V_C \Delta M_o + N_A V_C (2M_{o,in} - M_{out}) \exp\left(\frac{-\alpha t}{N_A V_C}\right)\end{aligned}$$

We now have an equation for the number of calcium ions inside the cell as a function of time. We want to calculate how long it will take for this number to reach a lethal number and kill the cell, which we write as $N_{lethal}$. We can then rewrite $N_{lethal} = N_a * V_c * M_{lethal}$, where $M_{lethal}$ is the lethal concentration. Making this substitution and solving for t produces the final equation:

$$t = \frac{-N_A V_C}{\alpha} \ln\left[\frac{M_{lethal}-\Delta M}{2M_{in}-M_{out}}\right], \text{ where } \alpha = \frac{A_p RT}{2\sqrt{3kTm}}$$

# VI. Acknowledgements

First and foremost, I would like to thank my mentor, Austin Sendek. I could not have done this project without him, for he truly put in the extra effort to make this work. It was a great pleasure to work with him over the past 11 months. Secondly, I would like to thank both Mr. Mark O'Connor and Dr. Jonathan Parsons for their interest and support throughout my entire project process, and also for fueling my passion for the sciences. Lastly, thank you to Mr. Sam Kotler, our Aβ pore expert, as well as Professor Rajiv Singh for their expertise and advice on our model.